\documentclass[11pt]{article}

\usepackage[english]{babel}

\usepackage[letterpaper,top=2cm,bottom=2cm,left=3cm,right=3cm,marginparwidth=1.75cm]{geometry}

\usepackage{amsmath}
\usepackage{float}
\usepackage{graphicx}
\usepackage{subcaption}
\usepackage[font={small, sf}]{caption}
\usepackage[colorlinks=true, allcolors=blue]{hyperref}

\title{\vspace{3cm}
A communication protocol based on \(NK\) boolean networks for coordinating collective action}
\author{Yori Ong \\ 
University of Groningen \\ 
yori.ong@gmail.com}
\date{}

\begin{document}
\maketitle

\begin{abstract}
In this paper, I describe a digital social communication protocol (Gridt) based on Kauffman's \(NK\) boolean networks. The main assertion is that a communication network with this topology supports infinitely scalable self-organization of collective action without requiring hierarchy or central control. The paper presents the functionality of this protocol and substantiates the following propositions about its function and implications:
\begin{itemize}
    \item Communication via \(NK\) boolean networks facilitates coordination on collective action games for any variable number of users, and justifies the assumption that the game's payoff structure is common knowledge;
    \item Use of this protocol increases its users' transfer empowerment, a form of intrinsic motivation that motivates coordinated action independent of the task or outcome;
    \item Communication via this network can be considered 'cheap talk' and benefits the strategy of players with aligned interests, but not of players with conflicting interests;
    \item Absence of significant barriers for its realization warrants a timely and continuing discussion on the ethics and implications of this technology;
    \item Full realization of the technology's potential calls for a free-to-use service with maximal transparency of design and associated economic incentives.
\end{itemize}

\end{abstract}

\section{Introduction}
The ideas proposed in this article result from addressing the following question: under which conditions can person A (Alice) do action \(X\) to intentionally cause person B (Bob) to do action \(Y\)? Alternatively phrased, it asks what it means for Alice to hold power over Bob's motivation for a certain action. This question is inspired by a state of the world that asks for behavior change interventions on large scales and in many forms. As stated, the question is underspecified. A reasonable answer would be that it depends on the specific action \(Y\) which Alice intends to elicit from Bob and the relation between the two. However, it is worth exploring where a rational approach to this question can lead while maintaining generality as much as possible. Therefore, I refrain from overspecifying the intended outcome \(Y\) or characteristics of Alice and Bob, but I will desire that a general solution is scalable, repeatable, and valid under common knowledge (i.e.: does not hinge on any person knowing less than others). 

Common knowledge is a pivotal concept in this discussion, referring to the state in which a everyone in a group knows \(\phi\), everyone knows that everyone knows \(\phi\), and so on. In the context of this article, I use the term to include 'sufficient common \(p\)-belief.' That is, everyone believes \(\phi\) with a certain probability \(p\), everyone \(p\)-believes that everyone \(p\)-believes \(\phi\), etc. The probability \(p\) can be considered sufficient to keep the optimal strategies of players in a game the same as they would be under certain knowledge of \(\phi\) \cite{Dalkiran2012}. This interpretation allows for common knowledge to be inferred from information that is public, and weakened by signs of others not knowing. Where common knowledge is mentioned in this article, a reader may interpret it as referring to a sufficiently justified common \(p\)-belief.

Alice doing \(X\) contributes to Bob doing \(Y\), if it leads to Bob expecting an increased utility from doing \(Y\) compared to not doing \(Y\). For Alice to influence Bob intentionally, she must know how the utility that Bob expects from his actions is affected by her action \(X\). For example, this is the case if Alice an Bob commonly know that when Alice does \(X\) (commands), Bob is sufficiently rewarded for doing \(Y\) (obey) or penalized for not doing \(Y\) (disobey). Unequal ability to dispense monetary rewards or threats of harm often provide two people with sufficient common knowledge to infer a simple game structure.

In reality, such clear hierarchical relations between any pair of people are rare. We are often uncertain about how the utilities of other people are affected by actions of themselves and others. In Bayesian game theory, variations in the correspondence between people's actions and their utility is referred to as their player type. In uncertain situations, Alice doing \(X\) might lead to Bob doing \(Y\), but it might also deter him from doing \(Y\) and turn out costly for her if she misjudges Bob's type. Intentionally influencing the actions of others would require knowledge their player type first, or cleverly dealing with uncertainty.

Dealing with uncertainty in another person's type is discussed by Steven Pinker \cite{Pinker2008} in an example of how indirect speech can be used when bribing a police officer. A person pulled over for speeding would not blatantly offer the officer a bribe to let them go, since doing so could lead to an even bigger penalty. Instead, the person hides a bank note in his ID papers and asks the officer if 'they see a way to make the situation easier for both'. A corrupt officer will take the bribe and let the driver go. An honest officer may still know what the diver is trying to do, but will have no proof of a bribery attempt and can still only issue a ticket. This indirect speech strategy avoids the worst outcome for the driver in case the officer is honest, but still results in a mutually beneficial outcome with a corrupt (or cooperative) officer.

Common knowledge between individuals of one's intent to influence the other is essentially a proposal to define their relation as 'influencer' and 'influencee'. Both ratifying and rejecting such a proposal can damage the social standing of either individual. The indirect speech strategy therefore intends to avoid this common knowledge, and taps into common knowledge of a higher level cooperation game. At this higher level, a cooperative receiver will be given a reason to act as suggested, without this affecting the relation between the individuals. A scalable and generally applicable strategy to evoke cooperation from the right player types, without the risk of negatively affecting any relations could bring great value for orchestrating behavior change on a societal scale.

I illustrate in this article that a communication protocol based on Kauffman's \(NK\) boolean networks \cite{Kauffman1969} provides such a strategy. These networks are characterized by \(N\) nodes with a corresponding boolean output from a function that takes exactly \(K\) directed inputs from other nodes. This logic results in dynamical systems that transition between ordered and chaotic behavior, dependent on the connectivity parameter \(K\). As a communication network, the semi-regular and directed properties of the underlying graph have unique implications for common knowledge and social influence among connected individuals. The rationale behind this claim is illustrated in figure 1. Imposing this model onto social systems sparks a range of theoretical, technological and societal questions. Section two of this paper conceptualizes this protocol as a digital technology and explains its functionality. The third section motivates the protocol's utility for coordination, distribution of social influence and its relation with intrinsic motivation. Finally, I discuss the protocol in relation to the collective behavior of \(NK\) networks of automata and its potential societal implications.

\begin{figure}[h!]
    \centering
    \begin{subfigure}{0.3\textwidth}
        \includegraphics[width=\linewidth]{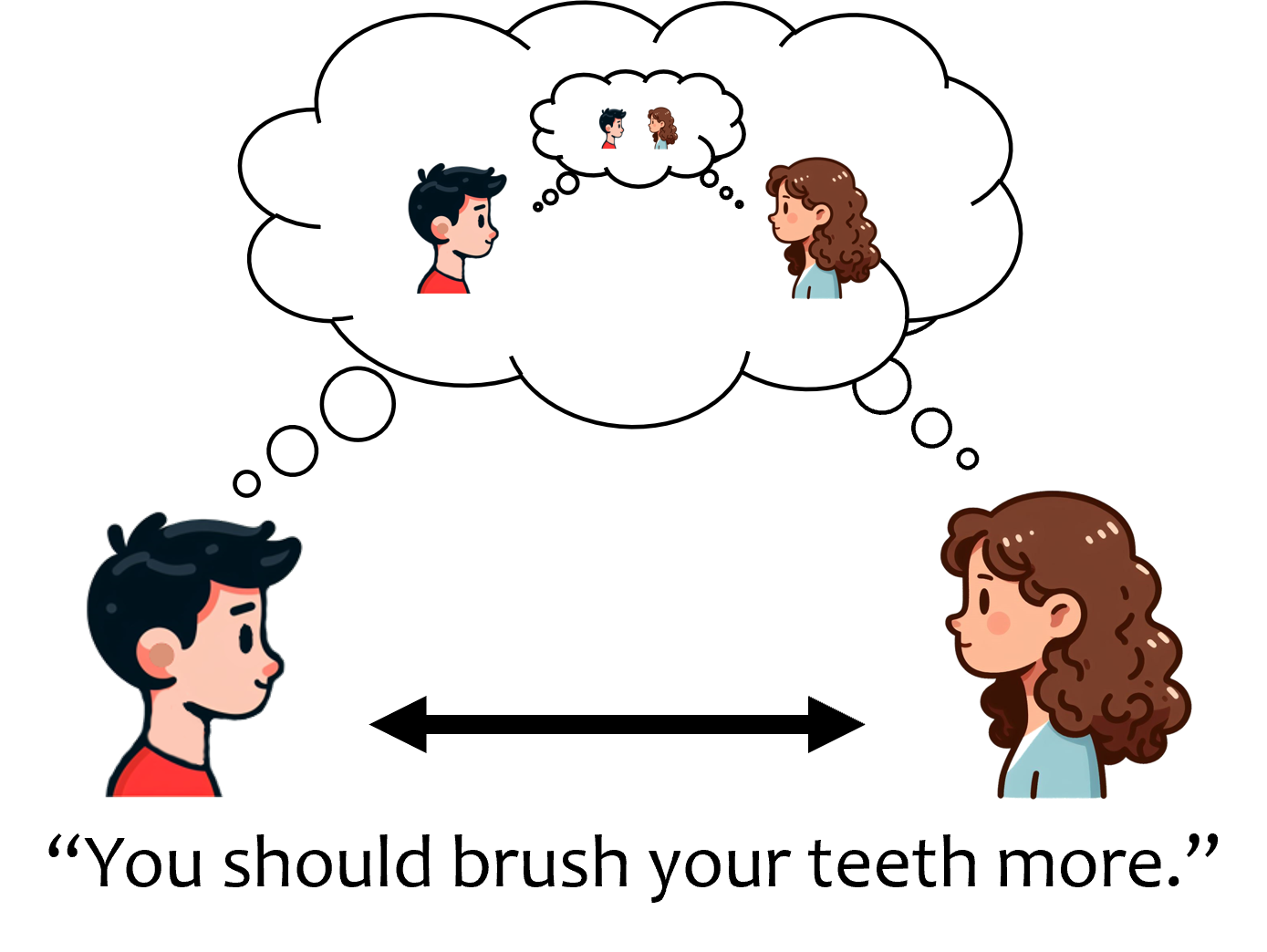}
        \label{fig:sub1}
        \vspace{-10pt}
        \caption{}
    \end{subfigure}
    \hspace{10pt}
    \begin{subfigure}{0.3\textwidth}
        \includegraphics[width=\linewidth]{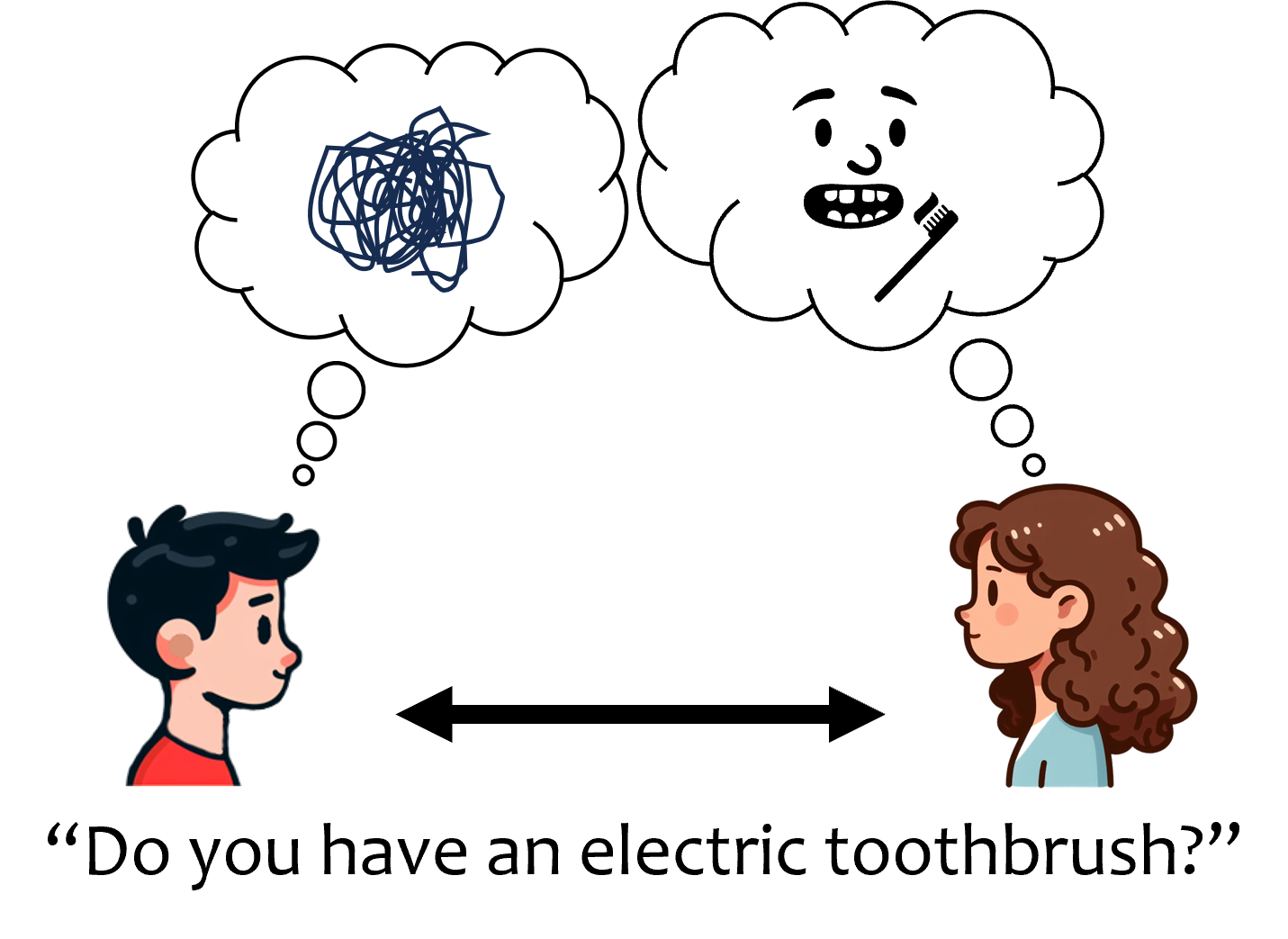}
        \label{fig:sub2}
        \vspace{-10pt}
        \caption{}
    \end{subfigure}
    \hspace{10pt}
    \begin{subfigure}{0.3\textwidth}
        \includegraphics[width=\linewidth]{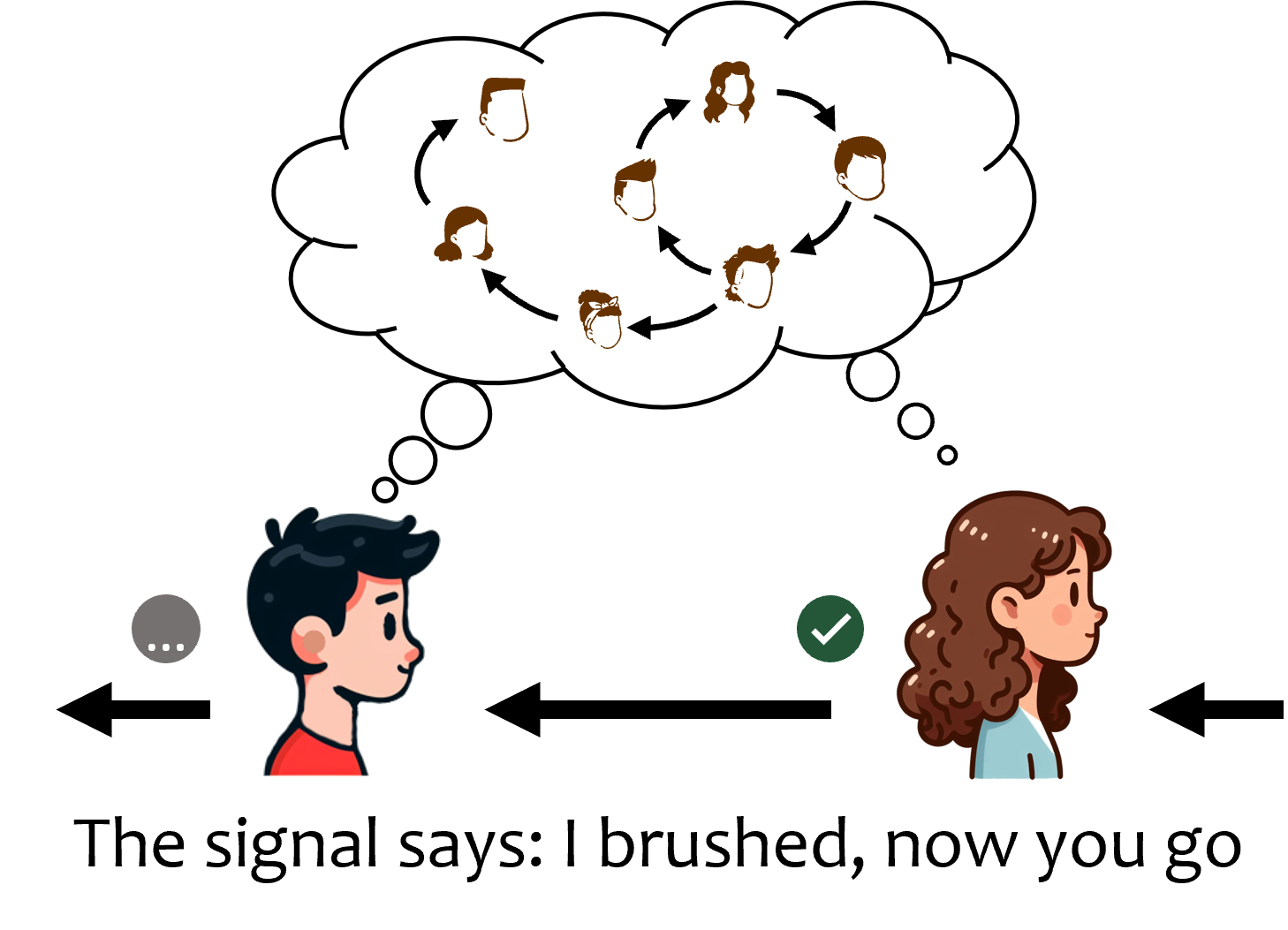}
        \label{fig:sub2}
        \vspace{-10pt}
        \caption{}
    \end{subfigure}
    \caption{(a) Direct speech is unambiguous, making Alice's attempt to influence Bob common knowledge to both. Bob may or may not take it well. (b) Using indirect speech avoids common knowledge of what Alice wants to achieve, but Bob may not get her hint, or wonder what she is implying. (c) Transmitting signals over a directed network avoids common knowledge of any information going from sender to receiver, but still allows for unambiguous communication. It does require that the connection and communication protocol is common knowledge to all in the network.}
\end{figure}

\section{Communication protocol}
This section conceptualizes the communication protocol on the \(NK\) boolean network as a digital technology. The terms user and player are both used to refer to a person using the protocol, depending on the technological or theoretical context of the discussion.

Users of the protocol can make the autonomous decision to connect themselves to others in an \(NK\) network topology. These networks consist of \(N\) nodes/users, all of which receive input from exactly \(K\) other nodes through directed links. The primary input and output values of each node are \((0)\) or \((1)\). For the purpose of illustration, we choose \(K=4\), for which I provide argumentation later on. The following information associated with each user is continuously transmitted along the directed links to users who receive their input:
\begin{itemize}
    \item A username and profile;
    \item A boolean signal, set to \((0)\) by default, that the user can activate by changing it to \((1)\);
    \item An optional, free form message that the user can formulate, once their signal has been set to \((1)\);
\end{itemize}
A public information channel presents at least the following information to all users upon connecting to the network:
\begin{itemize}
    \item A specification of an action that users can take and confirm to others by setting their signal from \((0)\) to \((1)\);
    \item An implicit or explicit reward that can be expected by cooperating users who have activated their signal;
    \item The reset condition under which all \((1)\) signals are simultaneously set back to \((0)\), repeating the cycle.
\end{itemize}
This information remains public to all connected users. Outside of the network, this information is communicated publicly or privately. In addition, we may assign a private ID to each user in the network, which is privately known and communicated at the users discretion. The complete protocol itself is public information and its purpose and function can be summarized as follows:
\begin{enumerate}
    \item All users can deduce that they have common knowledge of the action on which they coordinate, the number of \(2^K\) possible signal combinations they can observe, their possible signal states \((0,1)\) and the fact that all their decisions are made autonomously.
    \item This common knowledge provides the basis for playing repeated coordination games, while the network lets users asynchronously communicate their decisions to each other.
    \item No information transmitted via this network converts to common knowledge between sender and receiver. (Alice seeing Bob does not imply that Bob sees Alice: users cannot see who receives their signal.)
    \item Any variable number of \(N\) users can self-organize into this network, while the relevant neighborhood properties of each node are conserved at all time.
\end{enumerate}

The directed network property gives users autonomous control over (re)wiring their incoming links, since doing this leaves the inputs of others unaffected. Therefore, users can either sample their links randomly, or link with specific others in the network after learning their IDs. The semi-regular property ensures that the number of inputs for all users and the average size of their output sets (users who receive their input) are both equal to \(K\). Adding users to the network at any time does not affect this condition. However, to maintain these conditions throughout the play of a coordination game, restrictions must apply to users who want to disconnect and rewire their links. Therefore, rewiring incoming links is restricted to users who have made their final choice by setting their signal to \((1)\). Users who opt to leave can be disconnected from the network as soon as the reset conditions of the game are met.

Communicating exclusively over directed links prevents any pair of users from seeing each other mutually and converting transmitted information to common knowledge. This also prevents the basis of common knowledge from being changed by internal communication. By eliminating the possibility of conversation, relations between users remain solely based on the purpose of coordinating their actions. Free expression is facilitated by the messages users can attach to their activated signal, but contradicting common knowledge will then result in a degree of self-inconsistency. This protects the game against disruption from within. Users who no longer wish to coordinate under the conditions of the network can still choose to leave and join or initiate other networks.

The importance of directed links does not mean that no information can ever be transmitted in the reverse direction. As long as this information does not disclose the number or identity of users in the outgoing set, the conditions for the protocol to function are upheld. For instance, a boolean signal activated anonymously 'from behind' would inform a user that their activity is seen by at least one active user. This information is valuable for coordination, but uninformative enough to uphold the basic conditions of the protocol. In the figures 2 and 3, the communication protocol is visualized with a basic user interface and diagram representing the information flow to and from a user.

\begin{figure}[H]
    \centering
    \begin{subfigure}{0.4\textwidth}
        \includegraphics[width=\linewidth]{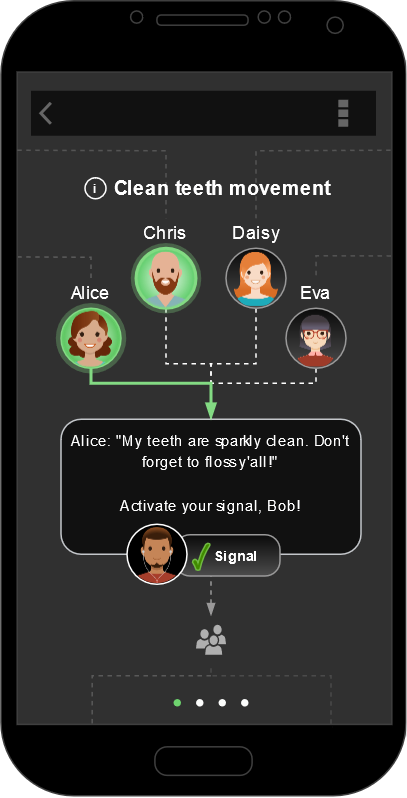}
        \label{fig:sub1}
    \end{subfigure}
    \hspace{20pt}
    \begin{subfigure}{0.4\textwidth}
        \includegraphics[width=\linewidth]{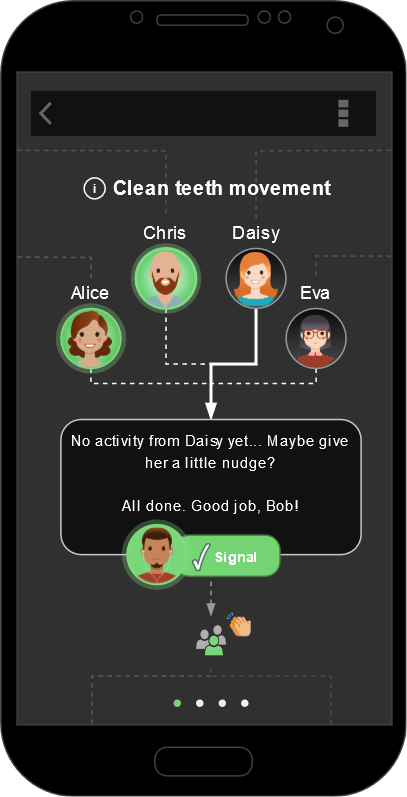}
        \label{fig:sub2}
    \end{subfigure}
    \caption{A basic application interface for communicating over the NK-boolean network, from the perspective of user Bob. Bob receives input information from Alice, Chris, Daisy and Eva. Left: Bob's signal is inactive. Right: Bob has activated his signal. Additionally he receives the signal that his activity is seen by at least one active user.}
\end{figure}

\begin{figure}[H]
    \centering
    \includegraphics[width=1.0\linewidth]{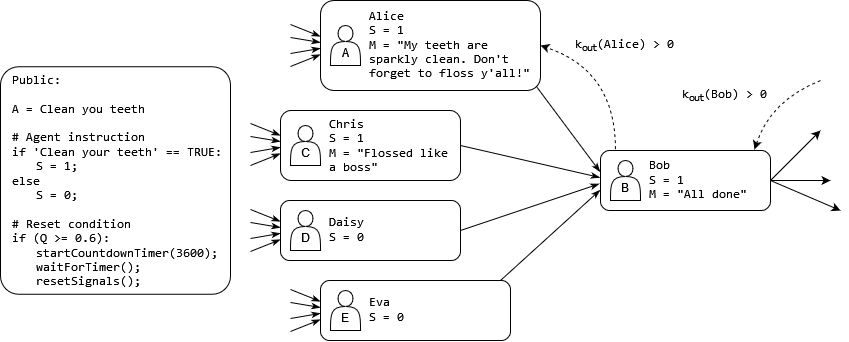}
    \caption{Diagram depicting the information received by and sent from user Bob, corresponding to the right screen of Figure 1.}
    \label{fig:enter-label}
\end{figure}

\section{Utility and relation with intrinsic motivation}
I have implied that this communication protocol is mainly useful for coordinating decisions in collective action games. These games are characterized by the highest utility being obtained by each player if all cooperate, while doing so is costly if the number of cooperators is too low. This results in non-cooperation always being a Nash equilibrium. Standard game theory offers no solution to such dilemmas, but in reality, common knowledge of the optimal solution is often a sufficient condition for people to coordinate successfully \cite{Thomas2014, Pinker2019}. Although the underpinning graph topology preserves common knowledge of the game structure at any scale, it does not make guarantee an instantaneous payoff from collective action in real life. Adoption of this communication protocol on a meaningful scale would thus depend on whether there is a task-independent utility associated with its use. In this section, I substantiate the claim that this protocol promotes cooperation, and that its use is driven by intrinsic 'empowerment' motivation.

The reasoning behind the assertion that the communication protocol promotes cooperation is based on users' inability to verify the truthfulness of the signals they receive. Signals indicate a user's cooperation with some collective action, which in reality may be true or not. The fact that users do not see each other mutually also removes possible social costs associated with lying. So, how is this helpful? Costless and unverifiable information exchanged between players in a game without limiting their strategies is referred to as 'cheap talk'. Assuming that the technology can indeed be used at no cost, and noting that it establishes the game structure for its players, the protocol provides them with a cheap talk channel. Theoretical and computational work has demonstrated that whether players transmit truthful information in equilibrium depends on the game's payoff structure \cite{Crawford1982, Cao2018, Farrell1996}. When players' interests fully align, communicating truthfully is a Nash equilibrium. In the case of fully conflicting interests, babbling (signaling with no correlation to the truth) is the only equilibrium. In intermediate cases, players can be expected to partially disclose private information. The utility of the communication protocol is therefore strongly biased toward players coordinating on actions based on common interest, since it provides no improved strategy for self-interested players.

Unlike in simulations, real collective action often comes with an uncertain and delayed reward, even if coordination succeeds. (Collective health behavior, climate adaptation and political protest are typical examples.) The use of a new technology to support coordination thus requires a form of motivation that is separable from its outcome, or intrinsic motivation. I specifically consider having causal influence on the actions of others to be this intrinsic motivator, which served as the introductory problem of this paper. The underlying network logic invites a computational perspective on this concept, even though it is focused on human decisions. In multi-agent reinforcement learning, the (potential) causal influence of one agent on another can be computed as the mutual information between an agent's action and another agent's action or observation. A proportional reward function for the influencing agent is referred to as transfer empowerment \cite{Salge2017}. Equipping artificial agents with this form of intrinsic motivation has been shown to promote emergent cooperation in multi-agent simulations \cite{Jaques2018, Heiden2020}.

Decentralized evaluation of transfer empowerment requires agents to have access to a model of other agents' inputs and processing functions. When internalized, such a model is equivalent to a basic 'theory of mind' mechanism: the capacity to assign mental states to others. Humans are capable of recursively nesting representations of others' mental states up to five orders \cite{Kinderman1998}. However, this also complicates predicting the influence our actions can have on others. Only under specific conditions can we construct a reliable model of others' mental state and transmit our influence to them. Connecting people in \(NK\) networks creates such a condition by both securing common knowledge of a coordination game and facilitating the necessary communication. This makes the evaluation of 'player-to-player' transfer empowerment a trivial matter: signaling cooperation will cause an observer to expect an increased payoff from their own cooperation. The fact that all players have the same information from which to derive their potential influence suggests that the basis for experiencing this empowerment is distributed evenly.

\subsection*{Connection to psychological theory}
The concept of computational transfer-empowerment as intrinsic motivation paints a consistent picture with validated psychological theory on the topic. Self Determination Theory, mainly developed by Ryan and Deci \cite{Deci1985} identifies autonomy, competence and relatedness as the three basic psychological needs for humans to thrive. I hypothesize that coordinating collective action over \(NK\) networks creates conditions that contribute positively to the user's three basic psychological needs. Competence refers to a person's need for the ability to meaningfully influence their own environment. If we consider other people to be part of this environment, transfer empowerment of one person to another is an aspect of competence. Coordination via the network can also produce a sense of joint competence in achieving goals that cannot be reached individually. The protocol also preserves autonomy of all users throughout, since the directed flow of information prevents any form of coercion. Furthermore, a sense of relatedness may be derived from the common knowledge of a shared goal between connected individuals.

Although digging deeper into the psychology of intrinsic motivation would result in a rich nuance, many psychological concepts are not operational from a computational perspective \cite{Oudeyer2009}. Since the operationalization of this communication protocol is highly feasible, unifying both perspectives becomes increasingly necessary. Various forms of intrinsic motivation have been formalized in (multi-agent) reinforcement learning and modeled in computational cognitive sciences. Some resulting definitions and architectures have a clear correspondence to fundamental aspects of human intelligence. Examples of such work include Schmidhuber's formal theory of creativity, fun, and intrinsic motivation \cite{Schmidhuber2010}, the Clarion cognitive architecture \cite{Sun2018} and intrinsic reward mechanisms derived from social influence \cite{Jaques2018}, empowerment \cite{Heiden2020} and theory of mind \cite{Oguntola2023}. The conceptualization of \(NK\) boolean networks for human communication holds potential for social experiments and thought experiments that help ground the social sciences in computational foundations. The influence of logical network architecture on the enhancement or suppression of certain human experiences can offer a glimpse into their computational nature. For example, preventing pairwise common knowledge in favor of collective common knowledge could impact experiences like trusting others, feeling disciplined, taking offense, or detecting hypocrisy. Conversely, computational models of cognition and intrinsic reward could be validated by their ability to account for aspects of real collective behavior on the network.

The implications of the proposed communication architecture on human behavior need to be understood from a computational frame in order to anticipate them. However, not formally understanding societal impact is typically not a barrier for technology to be realized, as long as the incentives for its development are sufficient. In fact, a common and informal language will be indispensable for the experiencing, disseminating and achieving acceptance of new possibilities. To reference the connection between the exact and social sciences, the communication protocol has been named 'Gridt', a blend of \textit{grid}, a network for distributing some property, and \textit{grit}, the trait of maintaining intrinsic motivation for goals with delayed payoff \cite{Duckworth2007}.

\section{The connectivity parameter \(K\)}
Building upon the preceding discussion on transfer empowerment, I will give a computational motivation for choosing the indegree \(K\) of all nodes on the network. The argument revolves around the fraction \(Q\) of cooperators, whose signal state is \((1)\). In a collective action game, the utility a player obtains is an increasing function of this fraction. Knowing the state variable \(Q\) means knowing whether a player's best choice is to cooperate or not. However, users of the Gridt network have no knowledge of \(Q\) or the number of users in the network by default, and are left to estimate these state variables based the information available to them.

We consider the situation in which Alice's signal is one of the \(K\) inputs seen by Bob. We denote Bob's estimate of \(Q\) by \(\tilde{Q}_B\), the state of Alice's signal by \(S_A = (0, 1)\), and the number of \((1)\) signals observed by Bob as \(\omega_B\). Using only \(K\) and \(\omega_B\) as input information, Bob's inferred probability density over \(\tilde{Q}_B\) would be a Beta distribution: \(\tilde{Q}_B \sim Beta(\omega_B+1 \:,\: K - \omega_B + 1)\). The influence of Alice activating her signal on Bob's estimate \(\tilde{Q}\) can then be quantified as a statistical distance between \(\tilde{Q}_B(K,\omega_B)\) and \(\tilde{Q}_B(K,\omega_B+1)\): 

\[I_A = D_{\mathrm{KL}} \Big[p(\tilde{Q}_B \,|\, S_A = 1, \omega_{B,\Bar{A}}) \:\Big|\Big|\: p(\tilde{Q}_B \,|\, S_A = 0, \omega_{B,\Bar{A}})\Big] \]

This expression measures change in Bob's estimate of the fraction of cooperators in the network caused by Alice's signal, while conditioning on Bob's other input signals (written as \(\omega_{B,\Bar{A}}\)). The use of the Kullback-Leibler divergence (\(D_{\mathrm{KL}}\)) as distance measure results in an expression proportional to the transfer empowerment from Alice to Bob. Since the optimal strategy in a collective action game is determined by the number of cooperators, Alice's signal has a potential influence on his action as well. The potential influence of an individual signal is inversely proportional to the \(K\), which is fixed for all users.

Additionally, we consider the probability \(P_{\emptyset}\) that a player's outgoing set is empty, under the assumption that the input set of each users is sampled uniformly at random from all others on the network. This probability as a function of \(N\) and \(K\) is given by \(P_{\emptyset} = (1-\frac{K}{N-1})^{N-1} \). On the domain \([K + 1, \infty)\), \(P_\emptyset\) is a monotonically increasing function of \(N\). In the limit \(N \rightarrow \infty\) this function reduces to \(P_{\emptyset} = e^{-K} \). Therefore, \( P_{\emptyset} < e^{-K} \), for a given \(K\).

The choice of \(K\) is therefore a trade-off between the potential influence of each signal on observer and the expected number of players in the outgoing set. Figure 3 displays an example of the case \(N=12\) and \(K=4\) with corresponding histogram of outdegree \(k_{out}\). The expected value of the influence of each signal \(I_a\) on each observer is computed for values of \(K\), where the distribution over \(\omega_{B,\Bar{A}}\) is taken to be binomial with \(p=0.5\). The resulting ballpark range for optimal \(K\) is the highlighted domain \([ 3,6 ]\), corresponding to the highest influence of an individual player's signal for which \(P_{\emptyset} < 0.05\).

We find that an increasing the connectivity parameter \(K\) increases the expected number of observers, but decreases the potential causal influence of each signal. The ballpark estimate \(K \in [3,6]\) shows interesting parallels with other works that theoretically and experimentally studied coordinating behavior in humans. Jiang et al. \cite{Jiang2021} studied the dependence of coordination failure rate in successive public goods games on the number of players \(M\). Their simulation showed low rates of failure in groups up to \(M=5\), which increased abruptly when the group size was increased to \(M=7\) and jumped again to certain failure for groups beyond \(M=19\) players. Social experiments with group sizes of 2 and 20 showed that all large groups failed to coordinate, while the outcomes of groups of 2 were more varied. The psychological interpretation given to these results was that smaller groups reduce the bystander effect. Centola et al. \cite{Centola2018} found empirically that in groups of 20 to 30 people, the influence of a committed minority of 25-31\% defying an existing convention is sufficient to overturn it. This was accurately predicted by their theoretical model, which converged on a critical mass of 25\% for large population sizes and many interaction rounds. For predicting collective behavior, these studies respectively modeled evolving strategies of agents based on realized payoffs and a contagion process among agents endowed with memory. These distinct approaches also seem to suggest that collective action takes off once roughly one in four cooperates. This motivates the choice of of \(K \sim 4\) on the Gridt network, which makes each individual signal represent \(\sim 25\%\) of the total input.

\begin{figure}[H]
  \centering
  \begin{subfigure}[b]{0.75\textwidth}
    \centering
    \includegraphics[width=\textwidth]{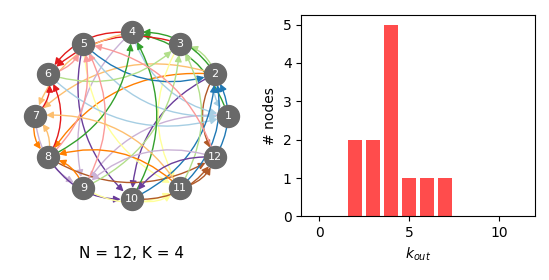}
    \caption{Example of a random \(NK\) network with \(N=12\), \(K=4\) and the outdegree histogram corresponding to the graph.}
    \label{fig:subfig1}
  \end{subfigure}
  \hfill
  \begin{subfigure}[b]{0.75\textwidth}
    \centering
    \includegraphics[width=\textwidth]{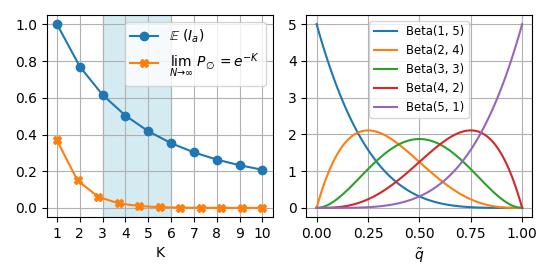}
    \caption{Left: Expected influence \(I_a\) and the probability \(P_{\emptyset}\) of \(k_{out} = 0\) as a function of \(K\). Right: Beta distributions for estimating \(Q\) for \(K=4\) and subsequent numbers of active signals.}
    \label{fig:subfig2}
  \end{subfigure}
  \caption{}
  \label{fig:combined}
\end{figure}

\subsection*{Dynamics of \(NK\) boolean networks}
Inherent to any discussion on \(NK\) boolean networks is their behavior as dynamical networks of automata. The original model defined by Stuart Kauffman \cite{Kauffman1969} was invented to study the regulation of gene expression. It consists of \(N\) automata, each with two possible output values \((0,1)\) produced by a random function of \(K\) inputs received from randomly chosen other automata. The behavior of the network is found to transition between an ordered and chaotic phase, depending on the connectivity parameter \(K\) and the bias of boolean functions toward returning the value \((1)\) \cite{Kauffman1984, Derrida1986}. In the ordered phase, the state of the network evolves in repeating patterns with short cycles (attractors) that are robust against external perturbations. In the chaotic phase, cycles of repetition are long and small perturbations propagate easily into new patterns of the system's behavior.

Kauffman's \(NK\) boolean network model is encountered mostly in studies of biological and physical systems, but makes no meaningful appearences in the social sciences. Clearly, humans are no simple automata, and the \(NK\) topology is not easily recognizable as a social interaction network. The Gridt protocol is also not intended as a model of an existing reality, but as an unexplored technological possibility to make social networks with this topology emerge. To interpret the meaning of ordered and chaotic behavior in this context, we first note that online communications can indeed causally influence offline behavior \cite{Centola2010, Althoff2017}. Properties of the network's collective dynamics can be determined with sufficient knowledge of how individuals respond to social influence, which depends on each player's utility function \cite{Lopez2008}. We may treat any other external motivators as perturbations to their response to social influence. The underlying graph can then be expected to promote collective human behavior that responds either predictably or capriciously to individual whims or external influences. The boundary between these phases, or the 'edge of chaos', is considered the optimal state in which complex adaptive systems balance between stability and adaptability \cite{Teuscher2022}. 

The Gridt protocol presents us with the interesting opportunity to control the connectivity parameter \(K\) in social networks, among other variables. Although the underlying graph adheres to the basic definition of the \(NK\) model, its implementation introduces important differences from the original model. These include a time-varying node and edge set, an expected skewness of the outdegree distribution, and the \((1) \rightarrow (0)\) transition being allowed only collectively. Previous theoretical and computational work is not directly applicable to the Gridt network, but does hint at what might be found. The annealed model of Derrida and Pomeau \cite{Derrida1986} shows the existence of a critical connectivity when the boolean functions and inputs of each node are randomly resampled at each time step. In boolean networks with scale-free directed and undirected topologies, the phase diagram forms a landscape determined by the network density, degree exponent and output bias \cite{Aldana2003, Fronczak2008}. The Gridt network presents the case of a fixed indegree and an outdegree that will likely tend toward a heavy-tailed distribution, as a result of network growth and individual users' connection preferences. Furthermore, a collective \((1) \rightarrow (0)\) transition imposes a periodicity on the system, which suggests that its dynamics could resemble cardiac tissue, in which the transition from tachycardia to fibrillation is a transition to spatiotemporal chaos \cite{Qu2011}. Tuning the connectivity parameter \(K\), recommendation algorithms for establishing links and collective action thresholds might thus be used to influence collective human behavior on various scales, without coercion or any traceable influence on people's individual decisions.

\section{Considerations for implementation}
What value for science and society can be expected from an actual digital implementation of the Gridt protocol? Who would set up such a digital infrastructure and control the parameters that determine the system's dynamics? And if it works as suggested, does it foster only constructive, or also destructive collective behavior? These are naturally arising questions, fueled by the ever growing impact of digital networks and algorithms on roughly 7 billion current users. The question whether it is a good idea to create this technology at all seems to present a false dilemma, if the incentives and resources to do so are widely available. In this final section, I briefly theorize which new strategies can be devised, whom they benefit, and which perspectives help us understand the potential consequences.

The Gridt protocol contributes to integrating the many disciplines and methods for studying collective human behavior. As a start, it bridges two main modeling approaches, based on contagion diffusion and an individual's utility, each of which has its respective strengths and weaknesses. Diffusion models replace the 'micro-mechanics' of human decision making with simple heuristics to accommodate the study of complex networks at larger scales. Utility-based models provide a general framework for the micro-mechanics of decision making, but make implicit assumptions of common knowledge that can often not be justified in uncontrolled social environments. Self-organization of people in \(NK\) networks accounts for the justification of common knowledge assumptions, a tractable game environment for each individual, and emergent complexity at large scales. Science-driven development of the technology can be motivated by new ways of measuring collective human behavior and further developing the theories that predict the system's behavior.

In addition to scientific advancement, the ability to orchestrate collective action toward a specific goal can contribute greatly to economic and social power.  So far, there has been no mention of the entities responsible for facilitating the digital infrastructure and controlling each network's parameters and public information channel. What are relevant decisions and payoffs for them and how do they affect the users' perception of a coordination game? We remind ourselves that users remain fully autonomous in their decision to join and cooperate, which they do based on the inference that the protocol and payoffs are commonly known to all. As mentioned in the introduction, the term common knowledge needs to be interpreted broadly to include a sufficiently certain common belief. To secure the protocol's function, I assert that this belief must also be justified and complete. This means that the algorithm implementation is public and verifiable, and no payoff-relevant information is kept hidden. Since confirmation of this belief cannot be obtained via the network itself, this information must be globally public. Therefore, successful coalitions between a technology provider, organizers and end users need transparent architectures for technology and business that best align the interests of those involved. Such interests include data-autonomy and economic fairness, which can respectively be secured by the use of decentralized web standards (e.g. Solid \cite{Sambra2016}) and purpose-driven business models (e.g. steward-ownership \cite{Purpose, Kavadis2023}.

Interestingly, the suggestions to deal ethically with privacy and profit did not result from weighing prescribed moral values. They instead seem to arise naturally as preferred strategies for optimizing individual utility by forming the most stable coalition, given possibilities of the Gridt protocol. One might suspect on this basis that a deeper logical connection exists between the \(NK\) network topology in social systems and the emergence of moral or prosocial behavior. However, such a connection does not mean that collective action is exclusively constructive. While many cooperation games yield greater payoffs as the coalition becomes larger (e.g. climate adaptation or lifestyle improvement), there are usually also players who benefit from collective non-cooperation. One can imagine that collective actions with economic or political consequences will always leave some who are negatively affected. The morality of self-organization is therefore not entirely indisputable and calls for a continuing investigation, which I leave as an exercise for the philosophers.

Despite the simple definition of \(NK\) boolean networks, their conceptualization as social communication networks gives rise to a complex adaptive system, whose characterization requires a transdisciplinary view. A discipline specific approach will quickly encounter violations of its basic assumptions. For example, analysing the possible outcomes of coalition formation games is the terrain of cooperative game theory. In this particular case, coalitions do not involve binding agreements between players, which is otherwise a basic assumption. Dealing with the resulting complexity, especially at larger scales, motivates the adoption of methods from other disciplines like statistical mechanics. The necessity of merging viewpoints arose previously in identifying aspects of individual utility and motivation. The expected emergence of this complex adaptive system is directly linked to the 'action' of the Gridt protocol. This action is to introduce a structure to agents that limits their communication to transmitting primary knowledge and establishing global common knowledge, while inhibiting recursive mentalizing. Approximate network regularity and simple information transmission seem to be recurring features in complex adaptive systems and may be recognized in flocking birds \cite{Ballerini2008} and schooling fish \cite{Hemelrijk2012}, but also in road traffic and party people forming a conga line.

\subsection*{Conclusion}
I have proposed that Gridt, a social communication protocol based on Kauffman's \(NK\) boolean networks, can trigger the emergence of self-organized collective action. Its directed and semi-regular topology secures common knowledge of a collective action game among players, and allows causal social influence to be transmitted between them without hierarchy or coercion. The high feasibility of implementing this protocol in the digital age calls for a timely, transdisciplinary investigation into its implications and ethics.

\bibliographystyle{unsrt}
{\footnotesize
\bibliography{Ref_Gridt1}}

\end{document}